# Lower solar atmosphere and magnetism at ultra-high spatial resolution

A White Paper for the Next Generation Solar Physics Mission

Remo Collet(1), Serena Criscuoli(2), Ilaria Ermolli(3), Damian Fabbian(4), Nuno Guerreiro(5), Margit Haberreiter(5), Courtney Peck(6), Tiago M. D. Pereira(7), Matthias Rempel(8), Sami K. Solanki(4), Sven Wedemeyer-Boehm(7)

(1)Stellar Astrophysics Centre, Department of Physics and Astronomy, Denmark; (2)National Solar Observatory, USA;( 3) INAF-Osservatorio Astronomico di Roma, Italy; (4) Max Plank Institute for Solar Physics, Germany; (5)Physikalisch-Meteorologisches Observatorium and World Radiation Center, Switzerland; (6)University of Colorado-LASP, USA; (7)Institute of Theoretical Astrophysics, University of Oslo, Norway; (8) High Altitude Observatory, USA

## Introduction

We present the scientific case for a future space-based telescope aimed at very high spatial and temporal resolution imaging of the solar photosphere and chromosphere. Previous missions (e.g., Hinode, Sunrise) have demonstrated the power of observing the solar photosphere and chromosphere at high spatial resolution without contamination from Earth's atmosphere. We argue here that increased spatial resolution (from currently ~70 km to ~25 km in the future) and high temporal cadence of the observations will vastly improve our understanding of the physical processes controlling solar magnetism and its characteristic scales. This is particularly important as the Sun's magnetic field drives solar activity and can significantly influence the Sun-Earth system. At the same time a better knowledge of solar magnetism can greatly improve our understanding of other astrophysical objects.

## Basic science topics

The physics of the outer solar atmosphere and heliosphere is largely driven by magnetic fields (e.g. Wiegelmann 2014). Moreover, solar magnetic activity has an impact on the surrounding space environment, including the Earth's atmosphere. The Sun is the yardstick of stellar evolution, activity, and variability studies and how exoplanets are affected by their host stars. Therefore, a deeper understanding of the Sun has an enormous impact on most of astrophysics, including stellar and planetary astrophysics. Below, we list some of the fundamental science topics that the ultra-high spatial resolution provided by a space-based 3+ meter telescope observing the solar photosphere and chromosphere would help to address.

### Magneto-convection/dynamo processes

The processes that produce the magnetic field and cause the solar activity cycle are not fully understood and likely require dynamo action on both large (driven by differential rotation, helical flows) and small (chaotic turbulent flows) scales. Small-scale magnetic field is ubiquitous in the solar photosphere and MHD simulations suggest an independent origin through small-scale dynamo (SSD) processes (e.g. Rempel 2014). Modeling SSD action in the photosphere is challenging due to the combination of a high magnetic Reynolds number and small magnetic Prandtl number. Such a regime is difficult to reach in numerical simulations as well as in liquid metal lab-experiments, making the solar photosphere a unique plasma laboratory. Ultra-high resolution observations could shed light on turbulent induction processes, feedback of the magnetic forces on small-scale turbulence in intergranular lanes and the associated modulation of the outgoing solar radiation. Numerical simulations greatly exceeding the resolution of available observations (see figure) show that currently unresolved magnetic fields do have dynamic consequences and impact outgoing solar radiation (e.g. Vögler 2005; Criscuoli 2013). Comparative studies of quiet and active solar regions will quantify the potential interaction between large- and small-scale magnetism and the role of the two dynamo processes as the key players in long-term changes of solar activity.

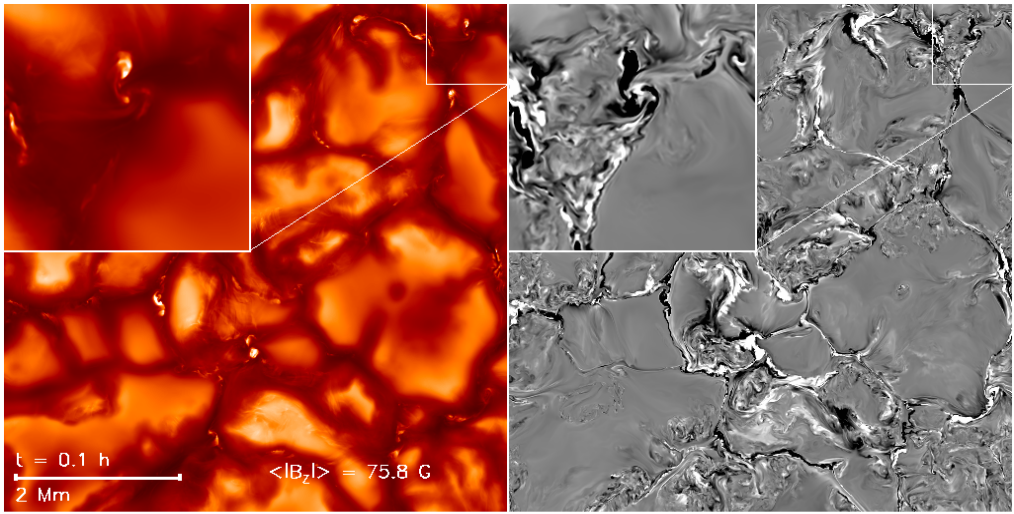

*High-resolution small-scale dynamo simulation of the solar photosphere with 4 km grid spacing. Left: bolometric intensity; right: vertical magnetic field at unity optical depth (saturated at +/- 400 G). Note the very small-scale structure of the mixed-polarity field, lost at the typical resolution of current solar telescopes.*

## Mass and energy flows: photosphere to corona coupling

Energy in the Sun is dissipated and converted on many spatial scales. The transport of mass and energy from the visible surface to the corona plays a critical role in driving solar activity, coronal heating, the solar wind, and the Sun-Earth connection, but the details of these processes are still poorly understood (e.g. Jess et al. 2015). It is believed that processes occurring at scales much smaller than currently observable are of enormous importance for the energy and mass balance. Observing the energy and mass flow at ultra-high spatial resolution is key to understanding not only the heating of the quiet corona, but also the buildup and impulsive phase of more dramatic events such as flares and possibly coronal mass ejections. The photosphere and chromosphere play a fundamental role in such processes.

A number of physical processes are not yet fully understood that affect the temperature-density stratification of the photosphere as well as the spectral energy distribution of the radiation originating there. For instance, the formation of active regions (sunspots and their penumbrae in particular), their evolution, dispersion, and interaction with the magnetism in the overlying layers, are still unclear (e.g. Cheung and Isobe 2014). Plasma dynamics and oscillations, their excitation and propagation to the higher layers of the atmosphere, the organization and interaction of multi-scale plasma motions, and their interaction with the ubiquitous magnetic fields, are important to understand fundamental physical processes occurring commonly under astrophysical conditions, but that can be studied with the necessary spatial, temporal, and spectral resolution only on the Sun.

Above the photosphere, the solar chromosphere is characterised by very fine structures as the plasma beta decreases. Heating processes such as reconnection and wave dissipation can occur on small spatial scales and on timescales of just a few seconds (e.g. Wedemeyer-Boehm et al. 2012). Understanding dynamic and intermittent structures on such spatial and temporal scales, as implied by simulations and current observations, is of fundamental importance because all mass and energy has to pass through this layer on its way into the corona. Observing the chromosphere is challenging because it is partially optically thin and most of its emission comes in the UV range, only accessible from space.

Observations of these layers at ultra-high resolution will uncover the processes acting there close to their fundamental scales. Such observations will in turn greatly profit from co-temporal observations probing the overlying corona in detail, although these need not be done by the same space mission.

# The mission: telescope and instruments

## Limitations of current and upcoming facilities

The upcoming DKIST will observe the solar surface with a 4-m aperture. While this represents a substantial improvement in terms of both spatial resolution and photon flux collection capability over any current solar telescope, image degradation and scattered light effects due to Earth's atmosphere will always be present to some degree - even when using adaptive optics and post-facto image reconstruction techniques. This limits the acquisition of uniform-quality data over time. Space-based instrumentation not only overcomes such limitations, but can also access wavelength ranges not accessible from the ground, particularly in the UV.

## Technical requirements

Achieving the science goals outlined above requires stable, seeing-free observations taken at a spatial resolution far superior (ideally 25 km or better) than the one presently achievable, over a wide field of view (FOV; enough to cover a medium-sized active region), and a broad wavelength range from the UV to the IR (e.g., Ly-alpha to 1.1 µm). High temporal resolution (namely, at a cadence of ~2 s) is also essential to clearly capture the rapid evolution of the observed quantities at the smallest scales.

The heart of such a mission would be a telescope with a 3-m or larger aperture, operating near its diffraction-limit in the visible, with high throughput and low scattered light. Such a size is feasible, being comparable with the 3.5-m telescope on ESA's Herschel mission. This telescope would feed instruments using a combination of novel and well-tested technologies, such as hyperspectral polarimetric imagers, or narrow- and broad-band (polarimetric-) imagers. Very large detectors or detector arrays will be needed to achieve high spatial resolution over a large FOV (e.g., 10k x 10k pixels, with a pixel size of 12.5 km on the Sun, covering 125 000 km). The large number of pixels and the high cadence imply an extremely high telemetry, making a geosynchronous orbit attractive.

The requirements above can be met by the following suite of instruments:
- A broad-band, wide FOV, multi-wavelength imager;
- A bi-dimensional tunable spectro-polarimeter, consisting of one or more channels;
- A (multi-)slit, spectro-polarimeter to cover multiple wavelength ranges;
- A hyperspectral polarimetric imager that gathers entire spectra at each spatial pixel.

The different instruments will observe simultaneously in the various spectral ranges (UV, visible and IR) that sample the solar atmosphere from the deep photosphere to the top of the chromosphere.

Although the diameter of such a space-based telescope will be smaller than the 4-m DKIST and planned EST ground-based telescopes, Hinode has impressively proven that even a smaller (0.5 m) telescope in space can provide breakthrough science that in many cases far surpasses what is delivered by ground-based telescopes with more than twice its diameter. The consistent quality and high stability provided by a space-based platform will be particularly important for time-series studies (e.g. waves and other dynamic phenomena), chromospheric and magnetic field physics (due to the low light levels, in particular when doing polarimetry and in the UV spectral range) and for catching important, but relatively rare phenomena such as large flares. We therefore expect that such a mission will produce many ground-breaking results of great relevance for all of solar physics, Sun-Earth relations and for astrophysics in general.

## Acknowledgements

The paper stems from fruitful discussions at the second meeting of the International Team "Toward New Models of Solar Spectral Irradiance based on 3D MHD simulations", ISSI, Bern, November 14-18, 2016. The authors are grateful to the International Space Science Institute for hosting and supporting the team.